\documentstyle[aps,12pt,manuscript]{revtex}
\begin{document}
\preprint{hep-th/9704199, INJE--TP--97--2 }
\def\overlay#1#2{\setbox0=\hbox{#1}\setbox1=\hbox to \wd0{\hss #2\hss}#1%
\hskip -2\wd0\copy1}

\title{Absorption of fixed scalar in scattering off 4D N=4 black holes}

\author{ H.W. Lee$^1$, Y.S. Myung$^1$ and  Jin Young Kim$^2$ }
\address{$^1$Department of Physics, Inje University, Kimhae 621--749, Korea\\
$^2$ Division of Basic Science, Dongseo University, Pusan 616--010, Korea}

\maketitle
\vskip 1.5in

\begin{abstract}
We  perform the perturbation analysis of the black holes in the 4D, N=4 supergravity.
Analysis around the black holes reveals a complicated mixing between
the dilaton and other fields (metric and two U(1) Maxwell fields).
It turns out that considering both s--wave ($l=0$) and higher momentum 
modes ($l \neq 0$), the dilaton as a fixed scalar is the only propagating mode 
with $P=Q, h_1=h_2=0$ and ${\cal F} =-{\cal G}= 2\varphi$.
We calculate the absorption cross-section for scattering of low frequency waves of  fixed scalar
and  U(1) Maxwell fields off the extremal black hole.

\end{abstract}

\newpage
Recently the absorption of fixed scalars into 4D extremal balck holes was considered 
in\cite{Kol,Kra}.  This is found to be suppressed compared with that of the free scalar. 
Actually the low energy absorption
cross section for  minimally coupled scalars turned out to be proportional to the area of the 
black hole horizon \cite{Dha,Das}. On the other hand, the fixed scalar cross section, as opposed to the 
Hawking decay, was 
found to vanish as $\omega^2$ in the classical limit
when $\omega \to 0$. The authors in\cite{Kol,Kra} used only part of the perturbing equations 
to decouple the
dilaton from other fields (metric and two U(1) Maxwell fluctuations).
However one has to consider all perturbing equations around the black holes
to find the consistent solution\cite{Cha}.

Holzhey and Wilczek have studied thoroughly the perturbations of the dilaton black hole with one
$U(1)$ charge\cite{Hol}. They started with fourteen equations. However there exist
only five (two metric, two U(1) Maxwell, and one dilaton) physical degrees of freedom.
The equations governing the perturbation of metric, U(1) Maxwell, and dilaton fields can be
reduced to five wave equations corresponding to five independent modes.
Actually these modes consist of various linear combinations of original functions parametrizing 
the perturbations and thus their direct physical meaning is not transparent.
We note that it is almost impossible to decouple the dilaton from other fields.
 Further the corresponding potentials are too
unwieldy to allow a useful description in closed form and analytical analysis.
Fortunately they found that the combined modes show the 
same qualitative behavior as the minimally coupled scalar. 
Since the field equation for a free scalar is remarkably simple, many authors consider it
as a spectator rather than real physical field in studying the black holes.

In this letter, we shall perform a complete analysis of the  perturbation
for the 4D, N=4 black holes with two U(1) charges\cite{Kal}.
This model with two U(1) charges provides us a prototype  for obtaining
the aborption cross-section of the decoupled dilaton (as a fixed scalar).
Also this analysis will be a cornerstone
for testing effective string models of the 5D black hole with fixed scalars\cite{Mal,Cal}.
The calculations of absorption and 
emission rates for relevant fields by the extremal and near-extremal black holes are very
important to compare them with the reults of D-branes.
Apart from the counting of states\cite{Vaf}, nowdays this is an important issue\cite{Das2}.

We  start with  the bosonic sector of the 4D, N=4 supergravity\cite{Kal}
\begin{equation}
S = \int d^4 x \sqrt{-g} 
   \left \{ R  - 2 (\nabla \phi)^2  - e^{-2 \phi} F^2 - e^{2 \phi}G^2 \right \}
\label{action}
\end{equation}
with the MTW conventions\cite{Mtw}.
Here the fields are metric $g_{\mu\nu}$,  dilaton $\phi$,  Maxwell fields  $F_{\mu\nu}$,
 and $G_{\mu\nu}$.  When $G_{\mu\nu}$ is absent, the above model reduces
to the dilaton black hole with $a=1$ in \cite{Hol}.
The equations of motion are given by

\begin{eqnarray}
&&R_{\mu\nu} - 2 \partial_\mu \phi \partial_\nu \phi -2 e^{-2\phi} F_{\mu\rho}F_{\nu}^{~\rho}
 +{1 \over 2} e^{-2\phi} F^2 g_{\mu\nu} -2 e^{2\phi} G_{\mu\rho}G_{\nu}^{~\rho}
 +{1 \over 2} e^{2\phi} G^2 g_{\mu\nu}=0,  
\label{eqR}\\
&& \nabla^2 \phi + {1 \over 2} e^{-2\phi} F^2 -{1 \over 2}e^{2 \phi} G^2 = 0,  \\
\label{eqD}
&&\nabla_\mu F^{\mu \nu} - 2 (\partial_\mu \phi) F^{\mu \nu} = 0,   
\label{eqF} \\
&&\nabla_\mu G^{\mu \nu} + 2 (\partial_\mu \phi) G^{\mu \nu} = 0.
\label{eqG}
\end{eqnarray}

The static charged black hole solution to the above equations is given by 
the background metric  
\begin{equation}
ds^2 = - (H_1H_2)^{-1}dt^2+H_1H_2 (dr^2+ r^2 d \Omega^2_2)
\label{defg}
\end{equation}
and 
\begin{equation}
e^{2\bar \phi} = { H_2 \over H_1},~~~ \bar F = { 1 \over \sqrt 2}dH_1^{-1}\wedge dt,~~~
 \bar G = { 1 \over \sqrt 2}dH_2^{-1}\wedge dt
\label{deff}
\end{equation}
with two harmonic functions
\begin{equation}
H_1= 1 +{ \sqrt 2 Q \over r}, ~~~~ H_2= 1 +{ \sqrt 2 P \over r}.
\label{class}
\end{equation}
Here $Q$ and $P$  are two U(1) charges.
We note that the event horizon is located at $r_{EH}=0$.
The fixed scalars are defined as the special massless fields whose values
on the horizon of the extremal black hole are fixed by the U(1) charges.
Considering (\ref{deff}), $\lim_{r\to 0} e^{2 \bar \phi}= P/Q$.  
In the extremal limit
this ratio is one and thus the dilaton can be regarded as a fixed scalar.
Similarly, we have $\lim_{r \to 0} \bar F^2= -1/ 2Q^2$ and
$\lim_{r\to 0}\bar G^2= -1/ 2P^2$. Therefore these may be considered
as the fixed tensors. Before we proceed, let us count the  degrees of 
freedom.  We have two graviton,
four U(1) Maxwell, and one dilaton degrees of freedom in four dimensions. 
Here for simplicity we start with five  among seven perturbation fields around
the  black hole background  as\cite{Lee}
\begin{eqnarray}
&&F_{tr} = \bar F_{tr} + {\cal F}_{tr} = \bar F_{tr} [1 + {\cal F}(t,r,\theta,\phi)],
\label{ptrF} \\
&&G_{tr} = \bar G_{tr} + {\cal G}_{tr} = \bar G_{tr} [1 + {\cal G}(t,r,\theta,\phi)],
\label{otrG} \\    
&&\phi = \bar \phi + \varphi(t,r,\theta,\phi),  \label{ptrD}  \\  
&&g_{\mu\nu} = \bar g_{\mu\nu} + h_{\mu\nu}.
\label{ptrg}
\end{eqnarray}
Considering the gauge transformation 
($h_{\mu\nu} \to h'_{\mu\nu} = h_{\mu\nu} - \nabla_\mu
\xi_\nu - \nabla_\nu \xi_\mu$), we can make $h_{\mu\nu}$ into the 
diagonal form and
further set the angular part of $h_{\mu\nu}$ to be zero.
Then the metric perturbation is given by two small fields $h_1(t,r,\theta,\phi)$ and $h_2(t,r,\theta,\phi)$\cite{Kra}
\begin{equation}
h_{\mu\nu}= {\rm diag}(- (H_1H_2)^{-1} h_1, H_1H_2 h_2, 0, 0).
\label{ptrmetric}
\end{equation}

 One has to linearize (\ref{eqR})-(\ref{eqG}) in order to obtain the equations governing the perturbations as
\begin{eqnarray}
&&  \delta R_{\mu\nu} (h) - 2(\partial_\mu \bar \phi \partial_\nu \varphi+
\partial_\mu \varphi \partial_\nu \bar \phi)
+2 e^{-2 \bar \phi}(-2 \bar F_{\mu \rho} {\cal F}_\nu^{~ \rho}   
 +\bar F_{\mu \rho} \bar F_{\nu\alpha} h^{\rho \alpha} 
+2\bar F_{\mu \rho} \bar F_\nu^{~ \rho} \varphi) \nonumber  
 \\
&& + e^{-2 \bar \phi} \bar g_{\mu\nu}(\bar F_{\sigma \rho} {\cal F}^{\sigma\rho}   
-  \bar F_{\rho \sigma} \bar F_\alpha^{~\sigma} h^{\rho \alpha} 
-  \bar F^2 \varphi) + {1 \over 2}e^{-2 \bar \phi} \bar F^2 h_{\mu\nu}  \nonumber  \\
&&+2 e^{2 \bar \phi}(-2 \bar G_{\mu \rho} {\cal G}_\nu^{~ \rho}   
+ \bar G_{\mu \rho} \bar G_{\nu\alpha} h^{\rho \alpha} 
-2\bar G_{\mu \rho} \bar G_\nu^{~ \rho} \varphi) \nonumber  \\
&& + e^{2 \bar \phi} \bar g_{\mu\nu}(\bar G_{\sigma \rho} {\cal G}^{\sigma\rho}   
-  \bar G_{\rho \sigma} \bar G_\alpha^{~\sigma} h^{\rho \alpha} 
+  \bar G^2 \varphi) + {1 \over 2}e^{-2 \bar \phi} \bar G^2 h_{\mu\nu} = 0,
\label{linR}
\end{eqnarray}

\begin{eqnarray}
&&  \bar \nabla^2 \varphi
- h^{\mu\nu} \bar \nabla_\mu \bar \nabla_\nu \bar \phi   
- \bar g^{\mu\nu} \delta \Gamma^\rho_{\mu\nu} (h) \partial_\rho \bar \phi
 + e^{-2 \bar \phi}\bar F_{\mu \nu} {\cal F}^{\mu \nu}                
 - e^{-2 \bar \phi} \bar F_{\mu \nu}   \bar F_{\rho}^{~\nu} h^{\mu\rho} 
- e^{-2 \bar \phi} \bar F^2 \varphi  \nonumber  \\
&&~~~~~~~~~ - e^{2 \bar \phi}\bar G_{\mu \nu} {\cal G}^{\mu \nu}                
 + e^{2 \bar \phi} \bar G_{\mu \nu}   \bar G_{\rho}^{~\nu} h^{\mu\rho} 
- e^{2 \bar \phi} \bar G^2 \varphi = 0,
\label{linD} 
\end{eqnarray}

\begin{equation}
  ( \bar \nabla_\mu -2 \partial_\mu \bar \phi ) 
    ( {\cal F}^{\mu \nu} - \bar F_\alpha^{~~\nu} h^{\alpha \mu}
                             - \bar F^{\mu}_{~~\beta} h^{\beta \nu} )                                        
      + \bar F^{\mu \nu} (\delta \Gamma^\sigma_{\sigma \mu} (h)       
      -2 (\partial_\mu \varphi)) 
 = 0,  \label{linF}
\end{equation}

\begin{equation}
  ( \bar \nabla_\mu +2 \partial_\mu \bar \phi ) 
    ( {\cal G}^{\mu \nu} - \bar G_\alpha^{~~\nu} h^{\alpha \mu}
                             - \bar G^{\mu}_{~~\beta} h^{\beta \nu} )                                        
      + \bar G^{\mu \nu} (\delta \Gamma^\sigma_{\sigma \mu} (h)       
      +2 (\partial_\mu \varphi)) 
 = 0,  \label{linG}
\end{equation}
where 
\begin{eqnarray}
&&\delta R_{\mu\nu} (h) = -{1 \over 2} \bar \nabla_\mu \bar \nabla_\nu h^\rho_{~\rho}
 - {1 \over 2} \bar \nabla^\rho \bar \nabla_\rho h_{\mu\nu}   
 + {1 \over 2} \bar \nabla^\rho \bar \nabla_\nu h_{\rho\mu}   
 + {1 \over 2} \bar \nabla^\rho \bar \nabla_\mu h_{\nu\rho},   
\label{delR} \\   
&&\delta \Gamma^\rho_{\mu\nu} (h) = {1 \over 2} \bar g^{\rho\sigma} 
( \bar \nabla_\nu h_{\mu\sigma} + \bar \nabla_\mu h_{\nu\sigma} - \bar \nabla_\sigma h_{\mu\nu} ).
\label{delGam}
\end{eqnarray}
Solving (\ref{linF}) and (\ref{linG}), one can express two U(1) scalars 
(${\cal F}$ and ${\cal G}$) in terms of $\varphi, h_1, h_2$ as
\begin{equation}
2{\cal F} = h_1 + h_2  + 4 \varphi,~~~~ 2{\cal G} = h_1 + h_2  - 4 \varphi.
\label{eqsol1}
\end{equation}
This means that on-shell, ${\cal F}$ and ${\cal G}$ are no longer the independent modes. 
Also from (\ref{linR}) six off--diagonal elements are given by
\begin{eqnarray}
(t,r) : & {1 \over 2} \left[ \left( 
{H_1' \over H_1} + {H_2' \over H_2} \right ) + {2 \over r} \right ] 
\partial_t h_2 +
\left ({ H_1' \over H_1} -{H_2' \over H_2}\right ) \partial_t \varphi = 0, 
\label{eqtr}\\
(t,\theta) : & \partial_t \partial_\theta h_2 = 0, \label{eqtth}\\
(t,\phi) :   & \partial_t \partial_\phi h_2 = 0, \label{eqtf}\\
(r,\theta) : & { 1 \over 2}( \partial_r - \Gamma^\theta_{r\theta}) 
\partial_\theta h_1  + { 1 \over 2r} \partial_\theta (h_1 - h_2 ) -
\left( {H_1' \over H_1} - {H_2' \over H_2} \right ) 
\partial_\theta \varphi = 0,  \label{eqrth}\\ 
(r,\phi) :   & { 1 \over 2}( \partial_r - \Gamma^\phi_{r\phi}) 
\partial_\phi h_1 + {1 \over 2r} \partial_\phi (h_1 - h_2) -
\left( {H_1' \over H_1} - {H_2' \over H_2} \right ) 
\partial_\phi \varphi= 0,  \label{eqrf}\\ 
(\theta,\phi) : &  {1 \over 2}( \partial_\theta - \Gamma^\phi_{\phi\theta}) \partial_\phi ( h_1 + h_2)= 0, \label{eqthf}
\end{eqnarray}
where the prime $(\prime)$ means the differentiation with respect to $r$.
And four diagonal elements of (\ref{linR}) take the form
\begin{eqnarray}
&(t,t) :&~~ -(H_1H_2)^2 \partial^2_t h_2 + 
(\partial^2_r + {2 \over r} \partial_r) h_1 +
{ 1 \over r^2} ( \partial^2_\theta + \cot \theta \partial_\theta +
   {1 \over \sin^2 \theta} \partial^2_\phi ) h_1 \nonumber \\
&&~~-\left({ H_2' \over H_2} +{H_1' \over H_1}\right) {(h_1'-h_2') \over 2} +
\left({H_1'^2 \over H_1^2} + { H_2'^2 \over H_2^2}\right)(h_1-h_2) \nonumber \\
&&~~ +{ 2 Q^2 \over r^4 H_1^2}( h_2 + 2 \varphi -2 {\cal F}) 
+{ 2 P^2 \over r^4 H_2^2}( h_2 - 2 \varphi -2 {\cal G})=0,
\label{eqtt} \\ %\end{eqnarray}
&(r,r) :&~~ -(H_1H_2)^2 \partial^2_t h_2 + 
\partial^2_r h_1 - {2 \over r} \partial_r h_2 +
{ 1 \over r^2} ( \partial^2_\theta + \cot \theta \partial_\theta +
   {1 \over \sin^2 \theta} \partial^2_\phi ) h_2 + \nonumber \\
&&~~ -4\left({ H_1' \over H_1} -{H_2' \over H_2}\right) \varphi'
-\left({ H_1' \over H_1} +{H_2' \over H_2}\right) {(3 h_1'+h_2')\over 2} \nonumber \\
&&~~ +{ 2 Q^2 \over r^4 H_1^2}( h_1 + 2 \varphi -2 {\cal F}) 
+{ 2 P^2 \over r^4 H_2^2}( h_1 - 2 \varphi -2 {\cal G})=0,
\label{eqrr} \\ 
&(\theta,\theta) :&~~{1 \over r^2} \partial^2_\theta ( h_1 + h_2) -{2\over r^2} h_2  +
\left( {H_1'^2 \over H_1^2} + {H_2'^2 \over H_2^2}\right) h_2 
+\left [ \left({ H_1' \over H_1} +{H_2' \over H_2}\right) + {2 \over r} \right] {(h_1'-h_2') \over 2} \nonumber \\
&&~~ -{ 2 Q^2 \over r^4 H_1^2}( h_1 + h_2 + 2 \varphi -2 {\cal F}) 
-{ 2 P^2 \over r^4 H_2^2}( h_1 + h_2 - 2 \varphi -2 {\cal G})=0,
\label{eqthth} \\ 
&(\phi,\phi) :&~~ {1 \over r^2 \sin^2 \theta} \partial^2_\phi ( h_1 + h_2) -
{2 \over r^2} h_2 +
\left( {H_1'^2 \over H_1^2} + {H_2'^2 \over H_2^2}\right) h_2 \nonumber \\
&&~~ +{ \cot \theta \over r^2} \partial_\theta (h_1 + h_2) + 
\left[\left( {H_1' \over H_1} + {H_2' \over H_2} \right ) +
{2 \over r} \right ] { {h_1' - h_2'} \over 2} \nonumber \\
&&~~ -{ 2 Q^2 \over r^4 H_1^2}( h_1 + h_2 + 2 \varphi -2 {\cal F}) 
-{ 2 P^2  \over r^4 H_2^2}( h_1 + h_2 - 2 \varphi -2 {\cal G})=0.
\label{eqff} 
\end{eqnarray}
The dilaton equation (\ref{linD})  leads to
\begin{eqnarray}
&&-(H_1H_2)^2 \partial^2_t \varphi + \varphi'' + {2 \over r}  \varphi'
+ {1 \over r^2} ( \partial^2_\theta + \cot\theta \partial_\theta + 
{1 \over \sin^2\theta} \partial_\phi) \varphi \nonumber \\
&&+\left({ H_2'^2 \over H^2_2} -{H_1'^2 \over H^2_1}\right) 
{ h_2 \over 2}  
+\left({ H_2' \over H_2} -{H_1' \over H_1}\right) 
{h_1'-h_2' \over 4}  \nonumber \\
&&+{ Q^2 \over r^4 H_1^2}(h_1 + h_2 + 2 \varphi -2 {\cal F})
-{ P^2 \over r^4 H_2^2}(h_1 + h_2 - 2 \varphi -2 {\cal G})=0.
\label{eqdilaton}
\end{eqnarray}

Let us first consider the s--wave ($l=0$ partial wave).  This implies 
that the fluctuations are functions of $t$ and $r$ only.  In this case 
the relavant relation comes only from (\ref{eqtr}) among six off--diagonal 
elements.  
By integrating (\ref{eqtr}) over time, we can obtain one relation
\begin{equation}
\left[\left( {H_1' \over H_1} + {H_2' \over H_2} \right ) +
  {2 \over r} \right ] h_2 = - 2 \left( {H_1' \over H_1} -{H_2' \over H_2} 
   \right ) \varphi.
\label{constraint}
\end{equation}
Adding $(\theta,\theta)$ and $(\phi,\phi)$ component  equations leads 
to the other relation
\begin{equation}
\left[\left( {H_1' \over H_1} + {H_2' \over H_2} \right ) + 
{2 \over r} \right ] (h_1' -h_2') =  
    -2 \left[\left( {H_1'^2 \over H_1^2} + {H_2'^2 \over H_2^2} \right ) 
      -{2 \over r^2} \right] h_2 -
    4 \left( {H_1'^2 \over H_1^2} - {H_2'^2 \over H_2^2} \right ) \varphi .
\label{h1pmh2p}
\end{equation}
Note that (\ref{constraint}) and (\ref{h1pmh2p}) are valid only for different 
metric perturbations($h_1 \neq h_2$) and $l=0$ case. 
Also one finds the relation,
\begin{equation}
\left[ \left( {H_1' \over H_1} + {H_2' \over H_2} \right ) + {2 \over r} \right 
]
(h_1' + h_2') = -4 \left( {H_1' \over H_1} - {H_2' \over H_2} \right )
  \varphi',
\label{eqprimes}
\end{equation} 
by substracting (\ref{eqrr}) from (\ref{eqtt}).  And 
substracting (\ref{eqprimes}) 
from (\ref{h1pmh2p}) leads to
\begin{eqnarray}
&&\left[ \left( {H_1' \over H_1} + {H_2' \over H_2} \right ) + 
{2 \over r} \right ]h_2' = 
-2 \left( {H_1' \over H_1} - {H_2' \over H_2} \right )
\varphi' \nonumber \\
&&~~~~~~~~~~~~~~~~~~~~~~~~~~~~~~~~ 
+ \left[ \left( {{H_1'^2} \over {H_1^2}} + {{H_2'^2 } \over {H_2^2}} 
\right ) - {2 \over r^2} \right ] h_2 +
2 \left( {H_1'^2 \over H_1^2} - {H_2'^2 \over H_2^2} \right ) \varphi.
\label{h2p}
\end{eqnarray}
But (\ref{h2p}) turns out to be redandunt because it can be derived 
by differentiation of (\ref{constraint}).
Only for s--wave perturbation, we can obtain the 
dilaton equation by substituting (\ref{constraint}) and 
(\ref{h1pmh2p}) into (\ref{eqdilaton}):
\begin{equation}
\left[ r^{-2} \partial_r r^2 \partial_r - (H_1H_2)^2 \partial_t^2 -
{{4(P+Q)^2} \over {r^2(\sqrt 2 P + \sqrt 2 Q + 2r )^2}} \right ] \varphi =0,
\label{decoupled}
\end{equation}
which is exactly the same as (13) of Ref.\cite{Kra}.  However, we have to 
consider both $l=0$ and $l \neq 0$ modes\cite{Mal1} equations to obtain the correct 
dilaton equation.  

Considering higher angular momentum modes($l\neq 0$), all  equations 
(\ref{eqtr})--(\ref{eqff}) 
are satisfied simultaneously when
\begin{equation}
 P=Q,~~~~~ h_1=h_2=0,~~~~~~ {\cal F}=-{\cal G}. \label{eqsol}
\end{equation}
In order to obtain the radial equation, we separate the dilaton as
\begin{equation}
 \varphi(t,r,\theta, \phi)_{lm} =  e^{- i \omega t} \varphi(r) Y_{lm}(\theta,\phi).
\label{harmonic}
\end{equation}
Inserting (\ref{eqsol}) and (\ref{harmonic}) into (\ref{eqdilaton}), one finds  
the fixed scalar equation 
\begin{equation}
\varphi'' + {2 \over r} \varphi' + { (r+1)^4 \omega^2 \over r^4} \varphi - 
{{l(l+1)} \over r^2} \varphi -
{2 \over r^2 (r+1)^2} \varphi= 0, \label{eqvarf}
\end{equation}
where  the mass of the black hole is chosen
as $M=\sqrt 2 Q =\sqrt 2 P =1$. 
Considering (\ref{eqsol1}) and (\ref{eqsol}), the equations for U(1) fluctuations (${\cal F}, {\cal G}$)
are identical to  the fixed scalar equation (\ref{eqvarf}).
If the last term in (\ref{eqvarf}) is absent, 
it corresponds to the free scalar.  
Here we wish to distinguish (\ref{eqvarf}) from (\ref{decoupled}). 
Eq.(\ref{decoupled}) is consistent with (\ref{eqvarf}) when $P=Q$ and
$l=0$.  
The authors of Ref.\cite{Kra} used only the s--wave fluctuation equations  
to decouple the dilaton from the other fields
and thus found the equation (\ref{decoupled}). Actually we 
consider both $l=0$ and $l \neq 0$ modes  
to obtain (\ref{eqvarf}). Also the U(1) fluctuations are not 
considered in\cite{Kra}.

Further, we can recover Kol and Rajaraman's  result\cite{Kol} if we 
transform the location of the 
event horizon form $r_{EH}=0$ to $r'_{EH}=1$   by $r \to r'-1$.
 And then neglecting the prime, we have the new metric element
\begin{equation}
ds^2= - { (r-1)^2 \over r^2} dt^2 + { r^2 \over (r-1)^2}dr^2 + r^2 d\Omega^2.
\label{shiftmetric}
\end{equation}
Here we note that this extremal black hole do have the non-zero area ($A_H= 4 \pi$).
One finds the s--wave equation from (\ref{eqvarf}),
\begin{equation}
\varphi'' + {2 \over r-1} \varphi' + { r^4 \omega^2 \over (r-1)^4} \varphi - 
{2 \over r^2 (r-1)^2} \varphi= 0,
\label{newf}
\end{equation}
which is the same form as  (19) in Ref.\cite{Kol}. The authors in\cite{Kol} used only the
$U(1)$ Maxwell fluctuations to decouple the dilaton. Eventually this case is  reduced to
our case, although they did not use the metric fluctuations.
This is because one finds $h_1=h_2=0$ in our results.

One way of understanding a black hole is to find out how it reacts to external perturbation
(fixed scalar). We visualize the black hole as presenting an effective potential barrier (or
well) to the on--coming waves\cite{Kim}.
Here let us derive the s--wave absorption cross-section for the  scattering of 
 dilaton (and Maxwell fields) off the extremal black hole 
according to Kol and Rajaraman\cite{Kol}.  This can first be done by 
solving the equaion (\ref{newf})
in three region: the near region ($ 1 < r < 1 + \omega / \sqrt 2$),
the intermediate region ($ 1 + \omega / \sqrt 2 < r < 2^{1/4} / \sqrt \omega$) and 
the far region ($ r > 2^{1/4} / \sqrt \omega$). The intermediate region is necessary
for matching the solutions. One must have the ingoging waves at the event horizon
$(r \to 1, \zeta \to \infty)$
as $e^{ i(\zeta -\omega t)}$, where $\zeta= \omega r/(r-1)$.
In the near region, one can thus  express the solution in terms of 
Coulomb wave functions as $\varphi_{near}=  i F_1(\zeta) + G_1(\zeta)$. In the intermediate region
one has  $\varphi_{inter}=  A / z + B z^2$ with $z= \zeta / \omega$. Also we have
both the incident ($C$) and reflected ($D$) waves
( $r \varphi_{far}=  C F_0(\omega r) + D G_0(\omega r)$) in the far region.
According to the matching conditions, one finds $B / A = i\omega^3 / 3$ and
$D / C = \omega  (2B - A) / (A+B)$. Finally the absorption coefficient 
is given by the flux conservation $({\cal A}(\omega) + {\cal R}(\omega) =1)$ as
\begin{equation}
{\cal A}(\omega)= 1 - \left | { 1+ i{ D \over C} \over 1 - i{D \over C}} \right |^2
        = { 36 \omega^4 \over 9 + 9 \omega^2 + 18\omega^4 + \omega^6 +  4 \omega^8}.
\label{absorp}
\end{equation}
In the low frequency approximation of $\omega \ll 1$,
the scattering of the dilaton off the extremal black hole is dominated by
the $l=0$ ($s$-wave) term.  This leads to ${\cal A}(\omega) = 4 \omega^4$.
This means that the differential cross-section is independent of $\theta$,
just as it was in the classical case.
The corresponding $s$-wave cross-section in four dimensions is then given by
the optical theorem,
\begin{equation}
\sigma_{s}^{fixed} = { \pi {\cal A}  \over \omega^2} = 4 \pi \omega^2.
\label{crossfix}
\end{equation}
On the other hand the free scalar cross-section is proportional to the area of 
the horizon as\cite{Das}
\begin{equation}
\sigma_{s}^{free} = A_H = 4 \pi \label{crossfree}
\end{equation}
for the low frequency scattering ($l=0$, $\omega \rightarrow 0$). This 
cross--section is four times the geometric
cross-section. This larger effective size is characteristic of long-wavelength scattering;
in a sense, these waves feel their way around  the whole sphere $(S^2)$ on the event horizon.

In conclusion, we carried out a complete perturbation anaysis of the black holes in
the 4D, N=4 supergravity. It is shown that there is a complicated mixing between the dilaton
and other fields (metric and U(1) Maxwell fields). Here we consider 
both $l=0$ mode and $l \neq 0$ higher modes.  Then the dilaton is 
decoupled from the metric
and Maxwell fields when  $P=Q, h_1=h_2=0$ and ${\cal F} =-{\cal G}= 2\varphi$. The first means that 
one has a fixed scalar in the extremal limit. Also the second implies that we have no propagating
gravitons in the extremal black hole. Finally the U(1) Maxwell perturbations are identical to 
that of the dilaton, and thus have the
same absorption cross--section as in (\ref{crossfix}).  
This is found to vanish as (\ref{crossfix}) when $\omega \to 0$.
The fixed scalars are defined as the special massless fields whose values
on the horizon of the extremal black hole are fixed by the U(1) charges.
In this sense  two U(1) Maxwell fields belong to the fixed tensors. Here, these are reduced to
the fixed  scalars $( {\cal F}, {\cal G})$. Thus it is understood that the U(1) Maxwell fields 
 have the same aborption cross-section as that of the dilaton in (\ref{crossfix}).

\acknowledgments

This work was supported in part  by the Basic Science Research 
Institute Program,
Ministry of Education, Project Nos. BSRI--96--2413, BSRI--96--2441 and 
Inje Research and Scholarship Foundation.

\end{document}